\documentclass{an}
\usepackage{psfig}
\usepackage{times}
\usepackage{graphicx}
\begin{document}
\lhead[\thepage]{Ralph Neuh\"auser:
Infrared imaging search for companions in $\beta$ Pic and Tuc/HorA}
\rhead[Astron. Nachr./AN~{\bf XXX} (200X) X]{\thepage}
\headnote{Astron. Nachr./AN {\bf 32X} (200X) X, XXX--XXX}

\title{An infrared imaging search for low-mass companions to members of the
young nearby $\beta$ Pic and Tucana/Horologium associations \thanks{Based on
observations obtained on La Silla,
Chile, in ESO programs 65.L-0144(B), 66.D-0135, 66.C-0310(A), 67.C-0209(B),
67.C-0213(A), 68.C-0008(A), and 68.C-0009(A)} }

\author{R.~Neuh\"auser\inst{1,2}, E.W. Guenther\inst{3}, J. Alves\inst{4},
N. Hu\'elamo\inst{5}, Th. Ott\inst{2}, A. Eckart\inst{6} }
\institute{
Astrophysikalisches Institut, Universit\"at Jena, Schillerg\"asschen 2-3, D-07745 Jena, Germany \\
\and 
MPI f\"ur extraterrestrische Physik, Giessenbachstra\ss e 1, D-85740 Garching, Germany \\
\and
Th\"uringer Landessternwarte Tautenburg, Sternwarte 5, D-07778 Tautenburg, Germany \\
\and
European Southern Observatory, Karl-Schwarzschild-Stra\ss e 2, D-85748 Garching, Germany \\
\and
European Southern Observatory,  Alonso de Cordova 3107, Casilla 19001, Santiago, Chile \\
\and
I. Physikalisches Institut, Universit\"at zu K\"oln, Z\"ulpicher Strasse 77, D-50937 K\"oln, Germany
}

\date{Received April 2003; accepted July 2003}

\abstract{
We present deep high dynamic range infrared images of young nearby stars in the 
Tucana/Horologium and $\beta$ Pic associations, all $\sim 10$ to 35 Myrs young
and at $\sim 10$ to 60 pc distance. Such young nearby stars are well-suited for direct imaging
searches for brown dwarf and even planetary companions, because young sub-stellar
objects are still self-luminous due to contraction and accretion.
We performed our observations at the ESO 3.5m NTT with the normal infrared imaging
detector SofI and the MPE speckle camera Sharp-I.
Three arc sec north of GSC 8047-0232 in Horologium a promising brown dwarf companion
candidate is detected, which needs to be confirmed by proper motion and/or
spectroscopy. Several other faint companion candidates are already rejected
by second epoch imaging.
Among 21 stars observed in 
Tucana/Horologium,
there are not more than one to five brown dwarf companions
outside of 75 AU ($1.5^{\prime \prime}$ at 50 pc);
most certainly only $\le 5\%$ of the 
Tuc/HorA stars have brown
dwarf companions (13 to 78 Jupiter masses) outside of 75 AU.
For the first time, we can report an upper limit for the frequency of 
massive planets ($\sim 10~M_{jup}$) at wide separations ($\sim 100$ AU) 
using a meaningfull and homogeneous sample:
Of 11 stars observed sufficiently deep in $\beta$ Pic (12 Myrs),
not more than one has a massive planet outside of $\sim 100$ AU,
i.e. massive planets at large separations are rare ($\le 9~\%$).
}

\correspondence{rne@astro.uni-jena.de}

\maketitle

\section{Introduction: The target associations}

Extra-solar planets have not been detected directly, yet. Direct imaging detection 
is difficult, because of the problem of dynamic range: Planets are too faint and 
too close to their much brighter primary star. 
However, young planets are still contracting significantly, 
so that the fraction of self-luminosity compared to luminosity due to reflected 
light is much higher in young planets than in old planets; therefore, young 
planets are much brighter than old ones (e.g. Wuchterl \& Tscharnuter 2003).
Hence, young nearby stars should be promising targets for
direct imaging searches for sub-stellar companions, both 
brown dwarfs and giant planets.

\begin{table}
\begin{tabular}{lrlc}
\multicolumn{4}{c}{\bf Table 1. Our sample} \\ \hline
Star         & V     & Spectral & Parallaxe \\
             & [mag] & type     & $\pi$ [mas] \\ \hline
\multicolumn{4}{c}{Tuc/HorA (ZW00/Tor00): $\sim 35$ Myrs at $\sim 50$ pc} \\ \hline
CPD$-64^{\circ}120$  & 10.2 & K1 & \\
GSC 8047-0232        & 10.9 & K3 & \\
CoD$-53^{\circ}386$  & 11.0 & K3 & \\
HD 13183 (1)         &  8.7 & G5 & $19.93 \pm 0.79$ \\
GSC 8056-0482        & 12.1 & M3 & \\
SAO 232842  (2)      &  8.4 & G7 & \\ 
%
%
HD 177171 (3)&  5.2 & F7   & $19.07 \pm 0.79$ \\
HD 202947 (4)&  8.9 &K0+K2.5& $21.7 \pm 1.5$    \\
HD 207129 (5)&  5.6 & G0   & $63.95 \pm 0.78$  \\
HD 207575    &  7.2 & F6   & $22.18 \pm 0.80$  \\
HD 207964 (6)&  5.9 & F0+F1& $21.49 \pm 0.67$  \\
PPM 366328 (7)& 9.6 & K0   &                  \\
HD 224392    &  5.0 & A1   & $20.53 \pm 0.51$  \\
DS Tuc       &  8.0 & G6+G8& $21.6 \pm 1.3$    \\
HD 1466      &  7.5 & F9   &                   \\
HIP 1910     & 11.3 & M1   & $21.6 \pm 2.2$    \\
HIP 1993     & 11.5 & K7   & $27 \pm 24$       \\
HD 2884   (8)&  4.3 & B9   & $23.35 \pm 0.52$  \\ 
HD 2885 (6,8)&  4.5 & A2+A7& $19.0 \pm 4.4$    \\
HD 3003      &  5.1 & A0   & $21.52 \pm 0.49$  \\
HD 3221      &  9.6 & K4   & $21.8 \pm 1.0$    \\ \hline
\multicolumn{4}{c}{$\beta$ Pic (Z01): $\sim 12$ Myrs at $\sim 40$ pc} \\ \hline
HIP 23309    & 10.1 & K7   & $38.1 \pm 1.1$   \\
HD 35850     &  6.3 & F7   & $37.26 \pm 0.84$  \\
AO Men       & 10.1 & K3   & $26.0 \pm 1.0$    \\
HD 139084    &  8.1 & K0   & $24.2 \pm 1.1$    \\ 
HD 155555    &  6.9 & G5+K0+M4.5 & $31.83 \pm 0.74$  \\
PZ Tel       &  8.4 & K0   & $20.1 \pm 1.2$    \\
HR 7329      &  5.0 & A0+M8& $20.98 \pm 0.68$  \\
HD 181327    &  7.0 & F5.5 & $19.77 \pm 0.81$  \\
AT Mic (9)   & 10.3 & M4+M5& $97.8 \pm 4.7$   \\
AU Mic (9)   &  8.6 & M0   & $100.6 \pm 1.4$  \\
HD 199143    &  7.3 & F8+M2& $21.0 \pm 1.0$   \\ 
HD 358623    & 10.6 & K7+M3& (10)            \\ \hline
\end{tabular}
\\
Remarks: 
(1) Single-lined spectroscopic binary (Cutispoto et al. (2002).
(2) Secondary to SAO 232841 (8.7$^{\prime \prime}$ off), 
a double-lined binary (F8+K0) with weak Lithium (Tor00).
(3) Double-lined spectroscopic binary (ZW00).
(4) An eclipsing binary with a few day period according to the Hipparcos light 
curve solution and SB2 according to Cutispoto et al. (2002) with Li detected only
in the primary, but Ca H \& K emission in both components.
(5) Visual companion 3 mag fainter in V separated by about one arc min (WDS).
(6) Known sub-arc sec visual binary.
(7) Wide binary with CPD$-64^{\circ}4331$~B with $\Delta V=5.2$ mag (WDS).
(8) The binary HD 2885 and the triple HD 2884 form a quintuple.
(9) The binary AT Mic and the apparently single star AU Mic form a
very wide common proper motion pair (WDS).
(10) HD 358623 has the same proper motion as HD 199143 and is located nearby,
so that it most certainly has the same distance as HD 199143 (van den Ancker et al. 2000).
\end{table}

A few brown dwarfs in orbit around young stars were
confirmed so far by both proper motion and spectroscopy,
e.g. TWA-5 (Lowrance et al. 1999, Neuh\"auser et al. 2000b)
and HR 7329 (Lowrance et al. 2000, Guenther et al. 2001),
as well as two brown dwarf companions to the intermediate-age star HD 130948 
(Potter et al. 2002, Goto et al. 2002), which is 200 to 800 Myrs old, 
possibly a member of the UMa group (see Potter et al. 2002).

Whether there is a {\em brown dwarf desert}, i.e. few brown dwarf
companions, at large separations like it is found at small
separations with radial velocity variations, can be investigated
best around young stars with a few tens of Myrs of age, because their 
brown dwarf companions have already formed and are still bright.

In recent years, nearby associations within 100 pc were found, each of 
which with a number of young stars, which are the targets of our campaign:
The Horologium association (Torres et al. 2000; henceforth Tor00),
the Tucana accociation (Zuckerman \& Webb 2000, ZW00),
and the $\beta$ Pic group (Zuckerman et al. 2001b, Z01).
The young stars in Tucana and Horologium are located next to 
each other on the sky, have distances from 45 to 60 pc, and 
an age around 35 Myrs, i.e. are possibly just one single association
(Zuckerman et al. 2001a), which we call here Tuc/HorA.
The $\beta$ Pic moving group, though, may well be significantly younger,
namely only $\sim 12$ Myrs (Z01).

Very low-mass companions around any of those member stars 
can improve age estimates of the primary star 
and, hence, the whole association, namely by using 
pre-main sequence tracks, because very low-mass stars at this age range
(tens of Myrs) and all sub-stellar companions are above the main sequence,
while most primary stars are already on or very close to the main
sequence, so that age determination is difficult.

From all possible members of Tuc/HorA 
listed in table 1a in Tor00,
we selected for our observations the F-type stars with a Lithium equivalent 
width $W_{\lambda}$(Li)$\ge 100$ m\AA , 
the G-type stars with $W_{\lambda}$(Li)$\ge 200$ m\AA , 
the K-type stars with $W_{\lambda}$(Li)$\ge 300$ m\AA , 
and the M-type stars with $W_{\lambda}$(Li)$\ge 100$ m\AA~(see 
Neuh\"auser 1997 for a discussion of the Lithium
content per spectral type to be expected for pre-main sequence stars).
Of those ten targets, we observed those six stars listed in Table 1.

In the Tuc part of Tuc/HorA, 
we selected the most likely members from table 1A
in ZW00 which are listed as certain or possible nuclear members, adding others 
(from table 1A in ZW00) if they show either enough Lithium (criteria as above for 
Horologium stars)
or if they show IR excess emission (HD 181296 and HD 207129), but excluding three members 
(PZ Tel, HR 7329, and HD 181327), which are instead included in the $\beta$ Pic sub-sample.
We added HD 202947 to our sample, a K-type stars with weak Lithium, because it is listed as
eclipsing binary (of $\beta$ Lyr type) in Simbad and Hipparcos, which would be of particular 
interest for the determination of its stellar parameters. See our Table 1 for our full 
Tucana sample, 15 of 16 likely members were observed by us with SofI and Sharp, 
namely all but HD 202917.

For a complete listing of $\beta$ Pic moving group members, see Z01.
Three of those members (PZ Tel, HR 7329, and HD 181327) were previously listed 
as probable or possible members 
of Tuc/HorA 
(in table 1A of ZW00). 
Two other members are HD 199143 and HD 358623, presented as Capricornius association
of young nearby stars first by van den Ancker et al. (2000).
Of a total of 19 member systems listed in Z01, we observed twelve,
see Table 1 for the sub-sample observed by us.

In Table 1, we list all the stars observed by us 
with their V-band magnitude, spectral type, parallaxe,
and information about multiplicity (from Simbad).

In section 2, we present the observations and data reduction
and also list the resulting data.
Then, in section 3, we discuss detected companion candidates,
and in section 4, we investigate the dynamic range achieved
and discuss the relevance of non-detections on the
frequency of sub-stellar companions.

\section{Observation and data reduction}

We used two different IR imaging cameras, both at the 3.5m New Technology Telescope (NTT)
of the European Southern Observatory (ESO) on La Silla, Chile:
The MPE speckle camera SHARP-I (System for High Angular Resolution Pictures, Hofmann et al. 1992) 
with $256 \times 256$ pixel 
(scale 49.1 mas, determined from observing the Galactic Center in the same nights,
and confirmed by the two wide stellar binaries AT Mic A \& B 
and SAO 232842 \& SAO 232841) and the normal 1k $\times$ 1k IR imager Son of Isaac 
(SofI\footnote{see www.ls.eso.org/lasilla/Telescopes/NEWNTT/})
used in the {\em small SofI field} mode with a pixel scale of 144 mas,
the nominal pixel scale, confirmed by several wide stellar
binaries with known separations as measured by Hipparcos;
both pixel scales are measured with a precision of a few per cent.
We observed during seven campaigns listed in Table 2.

\begin{figure*}
\vbox{\psfig{figure=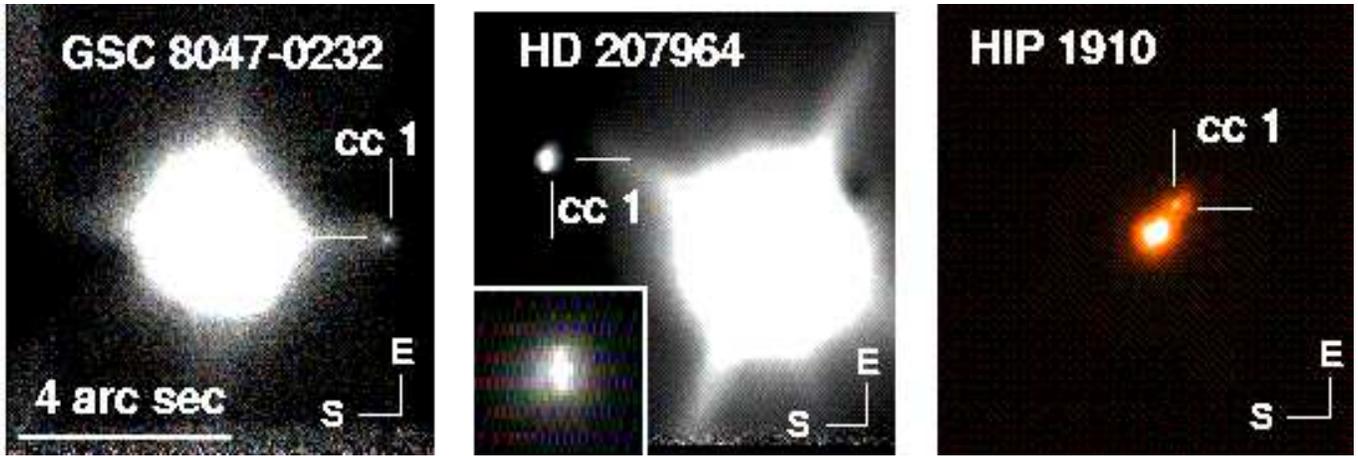,width=18cm,height=6cm}}
\caption{Our Sharp K-band images of GSC 8047-0232 (left), HD 207964 (middle), and 
HIP 1910 (right), where we have detected new companion candidates;
east up and south left, always $8^{\prime \prime} \times 8^{\prime \prime}$.
HD 207964 is a known sub-arc sec binary (WDS) somewhat elongated 
in our image (inlay in the central image, $1.7 ^{\prime \prime}$ aside),
the PSF of the companion candidate is consistent with being single.}
\end{figure*}

The FWHM in the final Sharp-I images ranges from $0.21 ^{\prime \prime}$
to $0.78 ^{\prime \prime}$ with the mean being $0.45 ^{\prime \prime}$;
and in the SofI images from $0.73 ^{\prime \prime}$
to $1.53 ^{\prime \prime}$ with the mean being $1.05 ^{\prime \prime}$
(see Table 3 for individual values).

\begin{table}
\begin{tabular}{lcc}
\multicolumn{3}{c}{\bf Table 2. Observing log} \\ \hline
Instrument & Dates (local)   & Program      \\ \hline
NTT/SofI   & 17-19 May 2000  & 65.L-0144(B) \\
NTT/SofI   & 07-08 Dec 2000  & 66.D-0135(A) \\
NTT/SofI   & 04-07 Mar 2001  & 66.C-0310(A) \\
NTT/Sharp  & 01-06 Jul 2001  & 67.C-0213(A) \\
NTT/SofI   & 08    Jul 2001  & 67.C-0209(B) \\ 
NTT/SofI   & 06-07 Dec 2001  & 68.C-0009(A) \\ 
NTT/SofI   & 08    Dec 2001  & 68.C-0008(A) \\ \hline
\end{tabular}
\end{table}

In addition to the science targets, we also observed several photometric standard stars
throughout each observing night (HR 8477, HR 7330, SAO 157131, HR 8278, HR 8658, and HR 6748). 
The nights 5/6 and 6/7 July 2001 were not photometric with some cirrus overhead,
so that those data cannot be used for absolute photometry.
Data reduction was performed in the normal way:
We removed bad pixels, subtracted the medium dark from all frames, then 
devided all science frames by a medium flat field, and then subtracted the sky.
Finally, we shifted and added the individual images up to a final image.
In Table 3, we list all individual observations with instrument used,
observing date, 
exposure time (on-source integration time without sky or overheads),
full-width at half-maximum (FWHM), filter used,
the observed (and previously known) magnitude of the primary,
as well as detection limits for undetected companions achieved in the observation
(in terms of detectable magnitude differences at different angular
and projected physical separations).

All detected companion candidates within (somewhat arbitrarily) 500 AU
are listed in Tables 4 \& 5 with separations, position angles, and magnitudes;
newly detected companion candidates, which are not yet rejected from
a 2nd epoch image, are shown in Fig. 1 (Sharp) and 2 (SofI).

\section{Results: Detected companions and candidates}

Let us first discuss the detected companions candidates (Table 5),
resolved known co-moving multiples (Table 4, top rows),
and then, in the next section, detection limits and the relevance 
of non-detections. All companion candidates found around DS Tuc,
PZ Tel, HD 139084, and AU Mic (listed in Table 4, bottom rows)
were found to be non-moving background objects by our own
2nd epoch observations.

{\bf GSC 8047-0232:}
The faint companion candidate cc 1 three arc sec north of GSC 8047-0232 
is the most promising in our data.
At $\sim 45$ pc distance, its absolute magnitude (M$_{V} = 11.1 \pm 0.3$ mag)
is consistent with a 25 to 30 Jupiter mass object (for 100 Myrs)
according to Chabier et al. (2000), or less massive for a younger age; 
at this absolute magnitude and age, it would be an early- to mid-L dwarf
(e.g. Leggett et al. 2002). With B.C.$_{K} = 3.3$ mag for early- to mid-L
(Leggett et al. 2002), its bolometric luminosity (at 60 pc) would be
$\log L_{bol}/L_{\odot} = -3.87 \pm 0.15$, corresponding to a mass of
$\sim 25$ Jupiters (at 35 Myrs) according to Burrows et al. (1997).
This companion candidate was detected independantly
by Chauvin et al. (2003) by coronographic JHK-band imaging with ADONIS 
at the ESO 3.6m on La Silla, having $J,H,K \simeq 16.2,15.2,14.9$ mag, 
i.e. consistent with our value K$=15.0 \pm 0.3$ mag. 
This red color is indicative of an L-type brown dwarf companion.
Follow-up spectroscopy can show whether it is a true companion
or a reddened background star.

{\bf PPM 366328:}
The faint coompanion candidates near PPM 366328, all at $H \simeq 19$ mag,
would be $\sim 5$ Jupiter mass objects (at 45 pc, 35 Myrs) according to
Burrows et al. (1997) with B.C.$_{K} = 3.3$ mag, but at a projected
physical separation of 400 to 500 AU, which is larger than expected for
planets, so that they are probably background objects, like
many of those in Table 4.
A 2nd epoch follow-up image can show whether any of the
companion candidates is co-moving with the primary star.

{\bf HD 207964:}
The faint companion candidate near HD 207964 at $K = 11.2 \pm 0.2$ mag,
would be a $\sim 40$ Jupiter mass object (at 46.5 pc, 35 Myrs) 
according to Burrows et al. (1997) with B.C.$_{K} = 3.3$ mag, 
i.e. possibly a brown dwarf companion. 

{\bf HIP 1910:}
The sub-arc sec companion candidate to HIP 1910 (separation of $639 \pm 13$ mas at 2001.5
with $K=9.1 \pm 0.1$ mag) was also detected by Chauvin et al. (2003) with $H \simeq 9.47$
and $K \simeq 9.44$ mag at a separation of $\sim 710$ mas. At the magnitude difference between 
the primary and the companion candidate ($\Delta K = 1.6 \pm 0.2$ mag), it would be an
early- to mid-M type companion, if bound. Then, its H-K color should be $\simeq 0.3$ mag. 
Because Chauvin et al. (2003) did not give errors to their magnitudes and separations
(nor their observing date), we cannot judge, yet, whether our results are consistent
and whether their H-K color for HIP 1910/cc 1 is consistent with early- to mid-M, 
nor whether the separation has changed between our and their observation.

\begin{table*}
\begin{tabular}{llccrc|ccc|cccc}
\multicolumn{12}{c}{\bf Table 3. Dynamic range achieved and limits on non-detections} \\ \hline
Star & Instr. & Obs. date & Expo. & \hspace{-0.5cm} FWHM & Band & \multicolumn{3}{c}{Primary magnitude} & \multicolumn{3}{c}{Magnitude limit (1)} \\
& & [UT] & [sec] & [mas] & & (2) & other & ref. & $0.5^{\prime \prime}$ & $1^{\prime \prime}$ & 100 AU \\ \hline
\multicolumn{12}{c}{Tuc/HorA (ZW00/Tor00): $\sim 35$ Myrs at $\sim 50$ pc, 13 Jup mass object has 17 mag in H or K} \\ \hline
CPD$-64^{\circ}120$ & Sharp & 04 Jul 01 & $1200 \times 0.5$ & 445 & K & 8.1 & 8.1 & (3) & 11.7 & 14.4 & 16.6 \\
GSC 8047-0232       & Sharp & 04 Jul 01 & $1200 \times 0.5$ & 236 & K & 8.7 & 8.4 & (4) & 12.3 & 15.0 & 17.2 \\
CoD$-53^{\circ}386$ & Sharp & 04 Jul 01 & $1200 \times 0.5$ & 385 & K & 8.6 & 8.5 & (4) & 11.7 & 14.2 & 15.9 \\
HD 13183            & Sharp & 04 Jul 01 & $1200 \times 0.5$ & 543 & K & 7.2 & 7.1 & (3) & 10.5 & 13.0 & 15.9 \\
GSC 8056-0482       & Sharp & 06 Jul 01 & $2400 \times 0.5$ & 577 & K & n/p & 7.5 & (4) & 10.9 & 13.2 & 15.4 \\
SAO 232842          & Sharp & 06 Jul 01 & $2400 \times 0.5$ & 511 & K & n/p & 6.5 & (4) & 10.0 & 12.6 & 15.2 \\ 
%
%
%
HD 177171    & Sharp & 07 Jul 01 & $2000 \times 0.3$ & 648 & K & n/p & 3.9 & (3) & 6.9 & 8.9  & 11.3 \\
%
%
HD 202947    & Sharp & 04 Jul 01 & $1200 \times 0.5$ & 580 & K & 6.8 & 6.6 & (4) &  9.8 & 11.6  & 15.9 \\
%
%
HD 202947    & SofI  & 09 Dec 01 & $500 \times 1.2$  & 864 & H & sat & 7.0 & (3) & sat & 11.6  & 13.8 \\
HD 207129    & Sharp & 05 Jul 01 & $1200 \times 0.5$ & 482 & K & 4.3 & 4.2 & (3) &  7.9 & 10.5  & off \\
HD 207575    & Sharp & 05 Jul 01 & $1200 \times 0.5$ & 447 & K & 6.0 & 6.2 & (4) &  9.9 & 12.8 & 15.6 \\
HD 207964    & Sharp & 05 Jul 01 & $1200 \times 0.5$ & 292 & K & 5.2 & 5.1 & (3) &  8.4 & 11.1  & 15.1 \\
PPM 366328   & SofI  & 20 May 00 & $460 \times 1.3$  & 1014& H & 7.8 & 7.7 & (3) & 10.5 & 11.8 & 15.5 \\
PPM 366328   & SofI  & 09 Dec 01 & $500 \times 1.2$  & 795 & H & 7.8 & 7.7 & (3) &  9.5 & 11.2 & 15.0 \\
HD 224392    & Sharp & 05 Jul 01 & $1200 \times 0.5$ & 490 & K & 4.9 & 5.1 & (3) & 8.8 & 11.6  & 13.2 \\
DS Tuc       & SofI  & 20 May 00 & $460 \times 1.3$  & 1170& H & sat & 6.4 & (3) & sat &  9.9 & 13.8  \\
DS Tuc       & SofI  & 09 Jul 01 & $400 \times 1.3$  & 1099& H & sat & 6.4 & (3) & sat & 11.0 & 13.3  \\ 
HD 1466      & Sharp & 05 Jul 01 & $1200 \times 0.5$ & 435 & K & 6.0 & 6.1 & (3) &  9.9 & 12.8 & 15.9 \\
HIP 1910     & Sharp & 05 Jul 01 & $1200 \times 0.5$ & 230 & K & 7.4 & 7.5 & (4) & 11.8& 13.8 & 17.1 \\
HIP 1993     & Sharp & 05 Jul 01 & $1200 \times 0.5$ & 352 & K & 7.6 & 8.3 & (3) & 12.3& 15.3 & 18.3 \\
HD 2884      & Sharp & 05 Jul 01 & $2000 \times 0.2$ & 356 & K & 4.4 & 4.5 & (3) & 8.4 &  11.2 & 13.9 \\
HD 2885      & Sharp & 05 Jul 01 & $2890 \times 0.2$ & 209 & K & 4.7 & 4.7 & (3) & 8.7 &  11.6 & 13.9 \\
HD 3003      & Sharp & 05 Jul 01 & $3000 \times 0.2$ & 389 & K & 5.0 & 5.0 & (3) & 9.0 &  11.6 & 14.3 \\
HD 3221      & Sharp & 04 Jul 01 & $1200 \times 0.5$ & 446 & K & 6.8 & 6.6 & (4) & 10.4 & 13.3 & 16.5 \\ \hline
\multicolumn{12}{c}{$\beta$ Pic (Z01): $\sim 12$ Myrs at $\sim 40$ pc, 13 Jup mass object has 15 mag in H or K}  \\ \hline
HIP 23309    & SofI  & 08 Dec 01 & $500 \times 1.2$  & 759 & H & sat & 7.1 & (3) & 11.6 & 12.9 & 16.7 \\
HD 35850     & SofI  & 09 Dec 00 & $400 \times 1.5$  & 896 & H & sat & 5.0 & (3) & sat & 13.4 & 18.4 \\ 
AO Men       & SofI  & 07 Dec 01 & $500 \times 1.2$  & 727 & H & sat & 7.0 & (5) & 12.6 & 14.1 & 15.9 \\
HD 139084    & SofI  & 18 May 00 & $460 \times 1.3$  & 1287& H & 6.2 & 6.2 & (3) & 8.0 & 8.1  & 10.8  \\
HD 139084    & SofI  & 09 Jul 01 & $400 \times 1.3$  & 1353& H & sat & 6.2 & (3) & 8.7 & 9.2 & 11.3 \\
HD 139084    & Sharp & 07 Jul 01 & $1200 \times 0.5$ & 783 & K & n/p & 6.1 & (3) & 11.9 & 14.8 & 18.2 \\
HD 155555    & Sharp & 06 Jul 01 & $1200 \times 0.5$ & 426 & K & n/p & 5.3 & (3) & 9.2  & 12.2 & 17.1 \\
PZ Tel       & SofI  & 18 May 00 & $460 \times 1.3$  & 1196& H & sat & 6.5 & (3) & sat  &  9.8 & 12.0 \\
PZ Tel       & SofI  & 09 Jul 01 & $400 \times 1.3$  & 994 & H & sat & 6.5 & (3) & sat  & 12.9 & 15.9 \\ 
HR 7329      & \multicolumn{11}{l}{see Lowrance et al. (2000), Guenther et al. (2001)} \\
HD 181327    & SofI  & 18 May 00 & $460 \times 1.3$  & 1085& H & sat & 6.0 & (6) & sat  & 11.9 & 14.4 \\
HD 181327    & Sharp & 06 Jul 01 & $2400 \times 0.5$ & 644 & K & n/p & 5.9 & (6) & 9.0  & 11.1 & 14.1 \\
AT Mic       & Sharp & 02 Jul 01 & $1200 \times 0.5$ & 534 & K & 4.9 & 4.9 & (5) & 8.2 & 10.5 & off  \\
AU Mic       & SofI  & 20 May 00 & $460 \times 1.3$  & 894 & H & sat & 5.1 & (3) & sat & 10.1 & 19.4 \\
AU Mic       & SofI  & 07 Dec 01 & $500 \times 1.2$  & 1534& H & sat & 5.1 & (3) & sat & 8.9 & 16.9 \\ 
HD 199143    & \multicolumn{11}{l}{see Jayawardhana \& Brandeker (2001), Chauvin et al. (2002), Neuh\"auser et al. (2002)} \\ 
HD 358623    & \multicolumn{11}{l}{see Jayawardhana \& Brandeker (2001), Chauvin et al. (2002), Neuh\"auser et al. (2002)} \\ \hline
\end{tabular}
\\
Remarks: (1) Magnitude limit for undetected but detectable point-like objects 
measured as $3 \sigma$ the flux in the bright star's PSF wing at that separation, 
converted to AU either with the known Hipparcos distance (Table 1) 
or the association's mean distance;
{\em sat} for saturated on our SofI image;
{\it off} means that this separation lies outside of the Sharp field-of-view.
(2) This work, typically $\pm 0.1$ mag, 
{\em n/p} for non-photometric conditions.
(3) Estimated from V-band magnitude and spectral type as listed in Table 1 (from Simbad)
using the colors given in Kenyon \& Hartmann (1995), $\pm$ a few tenth of mag,
because of errors in V, spectral type, color, and possible (unknown) IR excess.
(4) From DENIS (L. Cambr\'esy, priv. com.), $\pm$ a few tenth of mag near the DENIS saturation 
limit ($\sim 8$ mag), otherwise $\pm 0.1$ mag. 
(5) From 2MASS, $\pm$ a few tenth of mag near the 2MASS saturation limit ($\sim 5$ mag),
otherwise $\pm 0.1$ mag.
(6) Sylvester \& Mannings 2000.
\end{table*}

\begin{table*}
\begin{tabular}{lrrrrlcc}
\multicolumn{8}{c}{\bf Table 4. Companions and companion candidates detected twice} \\ \hline
Secondary (1) & $\Delta \alpha$ (2) & $\Delta \delta$ (2) & Proj. sep. & PA (3) & Comp.   & Date & Remarks \\ 
              & [mas]               & [mas]               & [AU] (2)   & [deg]  & H [mag] & ddmmyy & \\ \hline \hline
\multicolumn{8}{c}{Known wide companions (common proper-motion pairs):} \\ \hline 
PPM 366328 B    &$22980 \pm  2$ E&$9187  \pm 18$ S&$1113.7 \pm 0.8 $&$248.21 \pm 0.05$&$9.6 \pm 0.1$&20 05 00&(4)\\
                &$22978 \pm  2$ E&$9174  \pm  5$ S&$1113.4 \pm 0.3 $&$248.24 \pm 0.01$&$9.5 \pm 0.1$&09 12 01&(4)\\ \hline
DS Tuc B        &$1131  \pm 11$ W&$5191  \pm 31$ N&$246.0  \pm 1.5 $&$347.52 \pm 0.21$& saturated    &20 05 00&(4)\\ 
                &$1131  \pm 11$ W&$5191  \pm 28$ N&$246.0  \pm 1.4 $&$347.52 \pm 0.20$& saturated    &09 07 01&(4)\\ \hline \hline
\multicolumn{8}{c}{Rejected companion candidates (based on proper motions):} \\ \hline
DS Tuc A/cc 1   &$3511  \pm 17$ W&$8624  \pm 16$ S&$431.1  \pm 1.6 $&$202.15 \pm 0.15$&$16.3 \pm 0.3$&20 05 00 & (5) \\
                &$3589  \pm 20$ W&$8514  \pm 17$ S&$427.8  \pm 1.8 $&$202.86 \pm 0.18$& (n/p)          &09 07 01 & (5,6) \\ \hline
PZ Tel/cc 1     &$3168  \pm 53$ W&$6768  \pm 53$ N&$371.8  \pm 5.3 $&$334.37 \pm 0.62$&$15.9 \pm 0.3$&18 05 00 & \\
                &$3168  \pm 54$ W&$6912  \pm 53$ N&$391.4  \pm 5.3 $&$335.38 \pm 0.62$& (n/p)          &09 07 01 & (7) \\ \hline
HD 139084/cc 1  &$5760 \pm 300$ E&$1368 \pm 300$ N&$245    \pm 15  $&$163.0  \pm 3.8 $&$15.1 \pm 0.4$&18 05 00 & \\
                &$6192 \pm 300$ E&$1440 \pm 300$ N&$263    \pm 15  $&$166.9  \pm 3.6 $& (n/p)          &09 07 01 & (8) \\
HD 139084/cc 2  &$6927  \pm 12$ W&$2788  \pm 21$ N&$308.6  \pm 1.5 $&$291.92 \pm 0.21$&$15.0 \pm 0.3$&18 05 00 & \\
                &$6810  \pm 18$ W&$2966  \pm 30$ N&$306.9  \pm 1.8 $&$293.54 \pm 0.31$& (n/p)          &09 07 01 & (8) \\
HD 139084/cc 3  &$8219  \pm 27$ W&$4495  \pm 47$ S&$386.2  \pm 2.5 $&$241.57 \pm 0.38$&$15.1 \pm 0.3$&18 05 00 & \\
                &$8366  \pm 18$ W&$4477  \pm 30$ S&$392.1  \pm 1.8 $&$248.02 \pm 0.25$& (n/p)          &09 07 01 & (8) \\
HD 139084/cc 4  &$6193  \pm 11$ W&$8267  \pm 15$ N&$426.9  \pm 1.4 $&$323.16 \pm 0.11$&$ 9.7 \pm 0.2$&18 05 00 & \\
                &$6193  \pm 11$ W&$8322  \pm 15$ N&$428.7  \pm 1.3 $&$323.34 \pm 0.11$& (n/p)          &09 07 01 & (8)\\ 
HD 139084/cc 5  &$3610  \pm 12$ E&$9956  \pm 16$ S&$437.6  \pm 1.4 $&$160.07 \pm 0.10$&$11.9 \pm 0.2$&18 05 00 & \\
                &$3610  \pm 11$ E&$9819  \pm 14$ S&$432.3  \pm 1.3 $&$159.81 \pm 0.11$& (n/p)          &09 07 01 & (8)\\ \hline
AU Mic/cc 1     &$17136 \pm 52$ W&$12528 \pm 52$ S&$211.64 \pm 1.58$&$233.83 \pm 0.57$&$14.0 \pm 0.2$&20 05 00 & \\
                &$17828 \pm 34$ W&$11810 \pm 31$ S&$212.57 \pm 1.47$&$236.48 \pm 0.14$&$14.1 \pm 0.2$&07 12 01 & (9) \\
AU Mic/cc 2     &$25488 \pm 53$ E&$ 3312 \pm 53$ S&$256.25 \pm 1.59$&$ 97.40 \pm 0.58$&$14.7 \pm 0.2$&20 05 00 & \\
                &$25054 \pm 29$ E&$ 2616 \pm 51$ S&$250.40 \pm 1.52$&$ 95.96 \pm 0.14$&$15.0 \pm 0.2$&07 12 01 & (9) \\
AU Mic/cc 3     &$16704 \pm 53$ E&$25056 \pm 53$ N&$300.23 \pm 1.59$&$ 33.55 \pm 0.58$&$14.3 \pm 0.2$&20 05 00 & \\
                &$16139 \pm 26$ E&$25756 \pm 41$ N&$302.14 \pm 1.48$&$ 32.07 \pm 0.58$&$14.5 \pm 0.2$&07 12 01 & (9) \\
AU Mic/cc 4     &$12960 \pm 53$ E&$34488 \pm 53$ N&$366.23 \pm 1.58$&$ 20.60 \pm 0.58$&$13.2 \pm 0.2$&20 05 00 & \\
                &$12292 \pm 26$ E&$35088 \pm 38$ N&$369.57 \pm 1.47$&$ 19.31 \pm 0.58$&$13.3 \pm 0.2$&07 12 01 & (9) \\
AU Mic/cc 5     &$14544 \pm 53$ W&$38736 \pm 52$ N&$411.30 \pm 1.58$&$339.42 \pm 0.58$&$13.8 \pm 0.2$&20 05 00 & \\
                &$15107 \pm 25$ W&$39413 \pm 32$ N&$419.58 \pm 1.46$&$339.03 \pm 0.58$&$13.9 \pm 0.2$&07 12 01 & (9) \\
AU Mic/cc 6     &$26352 \pm 53$ W&$41256 \pm 52$ N&$486.62 \pm 1.58$&$327.43 \pm 0.58$&$13.5 \pm 0.2$&20 05 00 & \\
                &$26930 \pm 21$ W&$41920 \pm 30$ N&$495.28 \pm 1.45$&$327.28 \pm 0.45$&$13.7 \pm 0.2$&07 12 01 & (9) \\ \hline
\end{tabular}
\\
Remarks: (1) Companion candidates with already confirmed common proper motion 
(i.e. most certainly bound companions) are designated {\em B}, 
other companion candidates are called {\em cc}. We list only those objects 
detected within (somewhat arbitrarily) 500 AU around the primary target.
Those given with H-band magnitude are detected with SofI, those with K-band magnitude with Sharp.
(2) Separation errors for $\alpha$ and $\delta$ include $10\%$ error in pixel scale;
error for total separation (in AU) includes $15\%$ error for pixel scale and orientation
together plus the error in distance.
(3) Errors in position angles PA include $15\%$ error for pixel scale and orientation
together; PA is given as usual from North over East towards South.
(4) This is a previously known wide binary (WDS) confirmed here: The separation
has not changed or, putting it another way, the separation between the pair has
changed within the errors by much less than the known proper motion of the primary star.
(5) Separation and PA measured from the primary of the DS Tuc binary, i.e. the southern component.
(6,7,8,9) Background object because of significant change in separation ($\alpha$ or $\delta$
or both) and/or PA, namely according to the known proper motion of the primary,
but inconsistent with possible orbital motion.
From the offset changes between star and companion candidate(s), now found to be 
non-moving background object(s), we obtain the proper motion given in 
remarks (6,7,8,9),
always the correct direction and the correct order-of-magnitude 
(as in Hipparcos, see Perryman et al. 1997)
after only about one year epoch difference.
(6) We obtain $(\mu _{\alpha},\mu _{\delta}) =(65 \pm 26,-92 \pm 23)$ mas/yr
as proper motion, compared to Hipparcos: $(79.0 \pm 1.3,-67.1 \pm 1.1)$ mas/yr.
(7) We obtain $(\mu _{\alpha},\mu _{\delta}) =(0 \pm 75,-120 \pm 75)$ mas/yr
as proper motion, compared to Hipparcos: $(16.6 \pm 1.3,-83.58 \pm 0.87)$ mas/yr.
(8) We obtain $(\mu _{\alpha},\mu _{\delta}) =(-44 \pm 29,-77 \pm 41)$ mas/yr 
as proper motion, compared to Hipparcos: $(-52.9 \pm 1.2,-105.99 \pm 0.98)$ mas/yr.  
(9) We obtain $(\mu _{\alpha},\mu _{\delta}) =(365 \pm 174,-422 \pm 129)$ mas/yr
as proper motion, compared to Hipparcos: $(280.4 \pm 1.7,-360.09 \pm 0.79)$ mas/yr.
\end{table*}

\begin{figure}
\vbox{\psfig{figure=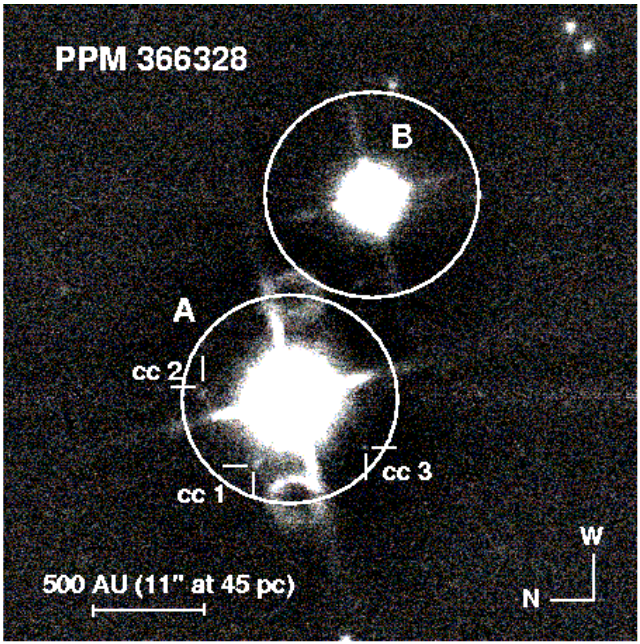,width=8.5cm,height=8.5cm}}
\caption{Our SofI H-band image of PPM 366328 with the stellar binary A and B
as well as three companion candidates within 500 AU around PPM 366328 A;
north left, west up, $60^{\prime \prime} \times 60^{\prime \prime}$.
The circles have $10^{\prime \prime}$ radii, i.e. 500 AU at the assumed
distance of Tuc/HorA (50 pc).}
\end{figure}

\begin{table*}
\begin{tabular}{lrrrrlcc}
\multicolumn{8}{c}{\bf Table 5. Companions and companion candidates detected once} \\ \hline
Secondary (1) & $\Delta \alpha$ (2) & $\Delta \delta$ (2) & Proj. sep. & PA (3) & Comp. mag. & Date & Remarks \\ 
              & [mas]               & [mas]               & [AU] (2)   & [deg]  & H or K (4) & ddmmyy & \\ \hline 
GSC 8047/cc 1   &$133   \pm 8 $ W&$3235  \pm 20$ N&$145.7  \pm 4.4 $&$357.65 \pm 0.18$&K=$15.0 \pm 0.3$& 04 07 01 & (5) \\
CoD$-53^{\circ}386$ B&$861.1  \pm 8.6$ E&$907.6 \pm 7.1$ S&$75.1 \pm 1.6$&$ 36.50 \pm 0.59$&K=$8.5 \pm 0.1$& 04 07 01 & (6) \\ 
HD 207964/cc 1  &$1674  \pm 12$ E&$4479  \pm 11$ S&$222.5  \pm 1.0 $&$159.51 \pm 0.21$&K=$11.2  \pm 0.2$ & 05 07 01& (7) \\
PPM 366328/cc 1 &$7992  \pm 52$ E&$ 4392 \pm 52$ N&$410.4  \pm 4.0 $&$ 60.21 \pm 0.51$&H=$19.3 \pm 0.4$&09 12 01 &(8)\\
PPM 366328/cc 2 &$6120  \pm 52$ E&$ 8280 \pm 52$ S&$463.3  \pm 4.0 $&$126.47 \pm 0.51$&H=$18.8 \pm 0.3$&09 12 01 &(8)\\
PPM 366328/cc 3 &$ 216  \pm 52$ W&$10584 \pm 52$ N&$476.4  \pm 4.0 $&$358.83 \pm 0.51$&H=$19.0 \pm 0.4$&09 12 01 &(8)\\
HIP 1910/cc 1   &$492   \pm 7 $ E&$ 407  \pm 11$ N&$ 29.55 \pm 2.3 $&$ 50.4  \pm 1.4 $&K=$9.1 \pm 0.2$  & 05 07 01 & (5) \\ 
HD 2885 B       &$519.7 \pm 9.9$ W&$26.1 \pm 7.6$ N&$ 27.4  \pm 4.4$&$272.88 \pm 0.98$&K=$5.4 \pm 0.1$  & 05 07 01 & (9) \\  \hline
\end{tabular}
\\
Remarks: (1) to (3) as in Table 4. (4) Companion magnitude given in H for SofI and in K for Sharp. 
(5) These companion candidates were also detected by Chauvin et al. (2003).
(6) Also called RST 47 A \& B (WDS), see Sect. 3.
(7) HD 207964 is a sub-arc sec binary (WDS), elongated in our image (see Fig. 1)
with an additional faint companion candidate (called here cc 1).
(8) The faint companion candidates are undetected in May 2000 due to limiting dynamic range
and sensitivity due to poor atmospheric conditions.
(9) Also called I 260 C \& D (WDS), a sub-arc sec binary widely separated 
from to the triple star HD 2884, see Sect. 3.
\end{table*}

\begin{figure*}
\vbox{\psfig{figure=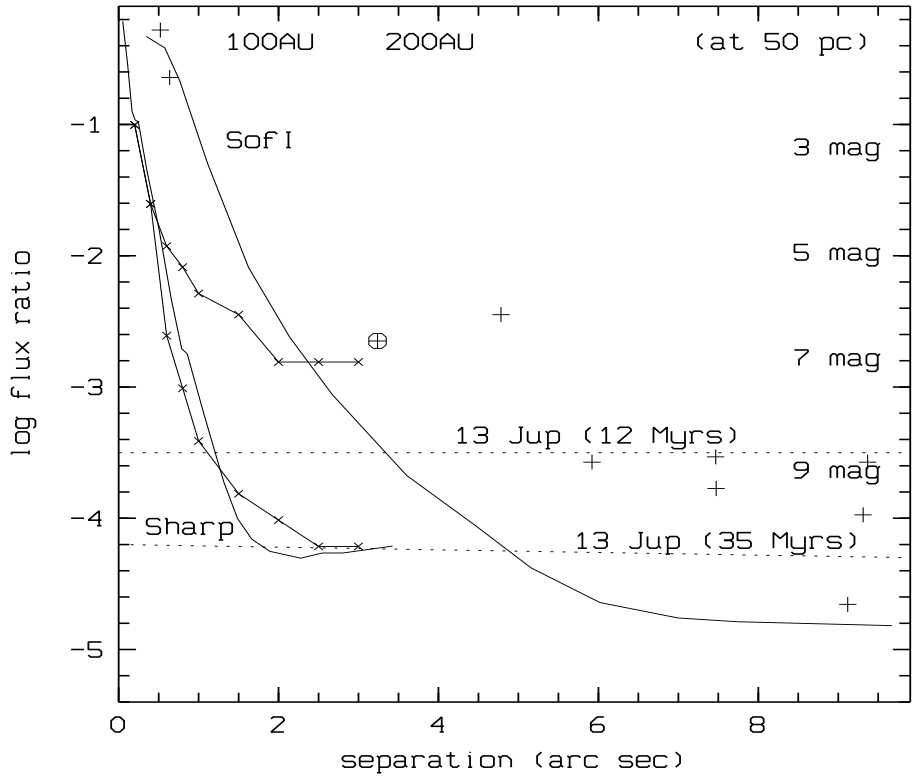,width=20cm,height=12cm,angle=270}}
\caption{Dynamic range achieved: Log of the flux ratio between
primary and $3 \sigma$ of the background limit versus separation
for both a typical Sharp image (lower full line) and a typical
SofI image (upper full line). We also indicate the expected 
flux ratios for 13 Jupiter mass companions at 12 (for $\beta$ Pic)
and 35 Myrs (for Tuc/HorA) as upper and lower broken lines,
respectivelly, plotted as flux ratio compared to a primary 
with the mean H-band magnitude in our sample (6.3 mag). 
Also shown are the observed companion candidates as crosses
with the most promising companion candidate GSC 8047/cc 1 circled. 
The two lines (with crosses inside) are the dynamic range achieved by Chauvin et al.
(2003) with ADONIS (with Sharp-II) at the ESO La Silla 3.6m without coronograph
(upper line) and with coronograph (lower line), taken from their figure 1.
Comparing those lines to other Sharp result, they are quite
identical within $0.5^{\prime \prime}$, and at wider separations,
our Sharp dynamic range is comparable to the ADONIS dyanmic range
when using the coronograph. Hence, as far as dynamic range is concerned,
Sharp-I at the NTT is as good as ADONIS at the 3.6m with coronograph.
The dynamic range achieved with the Sharp images shows
that we would have detected all brown dwarf companions, 
i.e.  all companions with mass above 13 Jupiters, at separations
outside of $1.0^{\prime \prime}$ at 12 Myrs (i.e. in $\beta$ Pic at 40 pc,
i.e. outside 40 AU) and outside of $1.5^{\prime \prime}$ at 35 Myrs
(i.e. in Tuc/HorA at 50 pc, i.e. outside 75 AU).
In the Sharp images of $\beta$ Pic stars, we could have detected massive
planets ($\sim 10~M_{jup}$) at 2 to $3^{\prime \prime}$ separations
(80 to 120 AU at 40 pc).
With SofI, 13 Jupiter mass companions are detectable only
outside of $\sim 4^{\prime \prime}$ (200 AU at 50 pc).}
\end{figure*}

{\bf HD 2884 and HD 2885:} 
HD 2884 ($\beta ^{1}$ Tuc) is a close 2.4$^{\prime \prime}$ binary (A and B) with large magnitude
difference (9.1 mag) according to the WDS with A being a spectroscopic binary (Aa and Ab),
so that HD 2884 forms an hierachical triple. The magnitude difference between A and B
is larger than our sensitivity limit (in a different band, though). 
The separation between the primary of the former (HD 2884 A) and HD 2885 
(also $\beta ^{2}$ Tuc or HD 2884 C) is 27$^{\prime \prime}$ with HD 2885 
being a close binary itself (C and D) with about half an arc sec 
separation (1.2 mag difference in WDS); the binary C and D (also called I 260, see WDS) 
was resolved by speckle imaging by Horch et al. (2000, 2001); they give $\Delta V = 1.2$ 
and $\Delta R = 1.2$ mag and list the binary as HD 2884 probably meaning C and D (i.e. HD 2885).
We have resolved this close binary, too (see Table 4).
Althogether, these objects form a quintuple.
I 260 C \& D have a separation of 0.58 to $0.59^{\prime \prime}$ at 1999.8 (Horch et al. 2000) 
the small change in separation (60 mas) and position angle ($4^{\circ}$) in a few
years indicates that this pair is most certainly a common proper motion pair ($5 \sigma$),
given the large proper motion of HD 2885 being $87.95 \pm 4.14$ and $-45.79 \pm 3.88$ mas/yr
(in $\alpha$ and $\delta$) according to 
Hipparcos (see Perryman et al. 1997).

{\bf CoD$-53^{\circ}386$:} This pair is RST 47 A \& B 
as listed in WDS, namely with $0.9^{\prime \prime}$ separation at a position 
angle of $312^{\circ}$ with $\Delta V = 0.1$ mag at epoch 1930 (Rossiter 1933).
The small change in separation in 71 years shows that the visual pair is a common proper
motion pair, i.e. most certainly bound, given the proper motion of the primary being
$38.2 \pm 3.5$ and $-23.0 \pm 3.3$ mas/yr (in $\alpha$ and $\delta$) according to 
Hipparcos (see Perryman et al. 1997).

\section{Discussion: Detection limits and brown dwarf companion frequency}

Let us now investigate the sensitivity limits determined for
the dynamic range achieved in the images:
The flux ratio is determined in all SofI and Sharp-I images as the 
$3 \sigma$ background noise level on $7 \times 7$ pixel boxes 
as approximate PSF areas and devided by the peak intensity.
We compare the observed dynamic ranges with expected flux ratios for possible
companions of different masses (calculated following Burrows et al. 1997)
next to a mean primary star (Fig. 3).

The MPE speckle camera Sharp-I clearly gives the best dynamic range.
In the Sharp images, we should have detected all sub-stellar companions
above $\sim 13$~M$_{jup}$, i.e. all brown dwarfs, outside of $\sim 1.0^{\prime \prime}$,
i.e. 40 AU (for the 12 Myrs young $\beta$ Pic members at 40 pc);
and in the SofI images, outside of $\sim 4^{\prime \prime}$,
any brown dwarf would have been detected.

From those numbers of non-detections, we can derive upper limits for the frequency of
brown dwarfs and massive planets in wide orbits around young stars:
For 12 Myrs young stars (the $\beta$ Pic members observed),
no massive planets of $\sim 10$M$_{jup}$ outside of 40 AU are detected
in a sample of six stars observed with Sharp (listed in Table 3 including 
HD 199143 and HD 358623 published in Neuh\"auser et al. 2002).
In the $\beta$ Pic group, we also detected 13 wide possibly sub-stellar
companions (listed in Table 4) around PZ Tel, HD 139084,
and AU Mic, all of which were rejected by our 2nd epoch observations,
i.e. all of them are unrelated background objects.
Hence, there are no brown dwarf companion candidates left in the $\beta$ Pic group
outside of $1.0^{\prime \prime}$ (40 AU) in eleven stars observed
either with Sharp or SofI.

As far as Tuc/HorA is concerned (35 Myrs at 50 pc), 
five possibly sub-stellar
companion candidates are not yet rejected nor confirmed, namely three around 
PPM 366328 A with 400 to 500 AU projected separation,
which are likely background stars, because they are very faint and widely separated,
one around GSC 8047-0232, which is probably a brown dwarf companion (JHK color ok),
and one around HD 207964, which may be sub-stellar (no color information).
Hence, among 21 stars observed in 
Tuc/HorA,
there are not more than one to five brown dwarf companions
outside of 75 AU ($1.5^{\prime \prime}$ at 50 AU);
most likely, just $\le 5\%$ of the Tuc/HorA stars have brown 
dwarf companions (13 to 78 Jupiter masses) outside of 75 AU (GSC 8047).
For more massive brown dwarf companions (35 to 78 Jupiter masses),
we can set even stronger contraints on the separation:
There is probably not more than one such brown dwarf among 21 stars 
observed detected outside of $0.5^{\prime \prime}$, i.e. 25 AU
(HD 207964). Hence, the frequency of wide brown dwarf companions is small.

Around some of our targets, we could have detected massive planets at
wide separations, see Table 3 for magnitudes expected for 13 Jup mass objects
and magnitude limits achieved at a separation of 100 AU (last column).
In Tuc/HorA, 
we can exclude planets with $\sim 10~M_{jup}$ outside
of $\sim 100$ AU only around GSC 8047, HIP 1910, and HIP 1993, too few
stars for a statistical anaylsis.
In the $\beta$ Pic group, however, where such planets would be brighter 
(because younger), we can exclude them around ten stars:
HIP 23309, HD 35850, AO Men, HD 139084, HD 155555, PZ Tel, and AU Mic
(this work) as well as HR 7329, HD 199143, and HD 358623 
(previous papers, see Table 3 for references).
For one additional star, AT Mic, we did not probe separations outside 
100 AU, because of its small distance and our 
limited Sharp field size\footnote{AU Mic, though, forming a very wide common proper motion 
pair with AT Mic, i.e. located at roughly the same distance, is probed outside of 100 AU, 
because it was observed with SofI, i.e. with a larger field; 
AT Mic, however, is located outside the field on the AU Mic SofI images}.
We cannot include HD 181327 in this statistic, because our 
sensitivity at $\sim 100$~AU separation (14.1 mag, table 3) is not 
deep enough to detect $\sim 10~M_{jup}$ mass objects (15 mag).
Hence, among 11 stars probed at around $\sim 100$ AU separations,
ten do not have wide massive planets, i.e. they are very rare ($\le 9~\%$).

Radial velocity surveys of nearby stars show that a significant
fraction ($\ge 8 \%$) have massive planets with orbital
radii substantially less than that of Jupiter
(Marcy \& Butler 2000, Udry et al. 2000, Butler et al. 2001).
Their close separation could be
explained, e.g., by in-situ formation, inward migration, or by a close encounter
with another (proto)planet, so that one planet has a very small, the other
a very large separation. Hence, one might expect a similar number of massive 
planets in very wide orbits as in very close-in orbits. 
Also, if the frequency of massive planets is constant in log mass (Zucker \& Mazeh 2002)
and increasing in log period (Armitage et al. 2002), than one should again
expect roughly as many massive planets in wide orbits as in close orbits.
By direct imaging, one can currently detect only planets in wide orbits.
Our upper limit for their frequency ($\le 9~\%$) is consistent with
the frequency of known Pegasi planets (few $\%$).
The non-detection of wide massive planets does not mean that encounters
of protoplanets are rare (due to limited statistics).
We note that none of the objects observed here is known to have a radial velocity 
planet candidate; young stars are difficult targets for radial velocity planet
searches because of the intrinsic activity, hence radial velocity scatter.

\acknowledgements
We would like to thank the NTT team with O. Hainaut, L. Vanci, and M. Billeres
for support during the SofI observations. We are gratefull to Klaus Bickert and
Rainer Sch\"odel for their help with the Sharp run.
Also, we are grateful to  Laurent Cambr\'esy, the DENIS consortium 
and its PI Nicolas Epchtein for providing us with unpublished data.
We also made use of the 2MASS public data releases.
We would like to thank Wolfgang Brandner and G\"unther Wuchterl
for many stimulating discussions.
RN and NH did most of their work for this project when they were at MPE; 
for support at that time, RN wishes to acknowledge the Bundesministerium 
f\"ur Bildung und Forschung grant number 50 OR 0003 distributed
through the Deutsche Zentrum f\"ur Luft- und Raumfahrt e.V.
We have made use of the Simbad database operated at the Observatoire Strassburg.

\refer
\aba

\rf{} Armitage P.J., Livio M., Lubow S.H., Pringle J.E., 2002, MNRAS 334, 248

\rf{} Baraffe I., Chabrier G., Allard F., Hauschildt P., 1998, A\&A 337, 403

\rf{} Burrows A., Marley M., Hubbard W. et al. 1997, ApJ 491, 856

\rf{} Butler R.P., Marcy G.W., Fischer D.A., et a., 2001, 
In: Penny A., Artymowicz P., Lagrange A.-M., Russell S. (Eds.)
ASP Conf. Ser., Planetary Systems in the Universe:
Observations, Formation and Evolution. PASP San Francisco, in press

\rf{} Chabrier G., Baraffe I., Allard F., Hauschildt P., 2000, ApJ 542, 464

\rf{} Chauvin G., Fusco T., Lagrange A.-M., et al., 2002, A\&A 394, 219

\rf{} Chauvin G., Dumas C., Beuzit J.L., et al., 2003, A\&A in press,
astro-ph/0304116

\rf{} Goto M., Kobayashi N., Terada H., et al., 2002, ApJ 567, L59

\rf{} Guenther E.W., Neuh\"auser R., Hu\'elamo N., Brandner W.,
Alves J., 2001, A\&A 365, 514

\rf{} Hofmann R., Blietz M., Duhoux P., Eckart A., Krabbe A., Rotaciuc V., 1992,
SHARP and FAST: NIR Speckle and Spectroscopy at the MPE. In: {\em Progress in Telescope
and Instrumentation Technologies}, Ulrich M.-H. (Ed.), ESO Conference and Workshop
Proc. 42, 617

\rf{} Horch E., Franz O.G., Ninkov Z., 2000, AJ 120, 2638

\rf{} Horch E., Ninkov Z., Franz O.G., 2001, AJ 121, 1583

\rf{} Jayawardhana R., Brandeker A., 2001, ApJ 561, L111 (JB01)

\rf{} Kenyon S., Hartmann L.W., 1995, ApJS 101, 117


\rf{} Leggett S.K., Golimowski D.A., Fan X., et al., 2002, ApJ 564, 452

\rf{} Lowrance P.J., McCarthy C., Becklin E.E. et al., 1999, ApJ 512, L69

\rf{} Lowrance P.J., Schneider G., Kirkpatrick J.D. et al., 2000, ApJ 541, L390

\rf{} Marcy G.W., Butler R.P., 2000, PASP, 112, 768

\rf{} Neuh\"auser R., 1997, Science 276, 1363

\rf{} Neuh\"auser R., Brandner W., Eckart A., et al., 2000a, A\&A 354, L9

\rf{} Neuh\"auser R., Guenther E.W., Brandner W., et al., 2000b, A\&A 360, L39

\rf{} Neuh\"auser R., Guenther E.W., Brandner W., et al., 2001,
{\it Direct imaging search for planetary companions next
to young nearby stars}. In: Montmerle T. \& Andre P. (Eds.)
From Darkness to Light. ASP Conf. Ser. 243, 723-728

\rf{} Neuh\"auser R., Guenther E.W., Mugrauer M., Ott T., Eckart A., 2002, A\&A 395, 877

\rf{} Perryman M.A.C., Lindegren L., Kovalevsky J., et al., 1997, A\&A 323, L49

\rf{} Potter D., Martin E.L., Cushing, M.C., Baudoz, P.,
Brandner W., Guyon O., Neuh\"auser R., 2002, ApJ 567, L133

\rf{} Rossiter R.A., 1933, Mem. RAS 65, 28

\rf{} Sylvester R.J., Mannings V., 2000, MNRAS 313, 73

\rf{} Torres C.A.O., Da Silva L., Quast G.R., de la Reza R., Jilinski E., 
2000, AJ 120, 1410 (Tor00)

\rf{} Udry S., Mayor, M., Naef D., Pepe F., Queloz D., Santos N.C., Burnet M.,
Confino B., Melo C., 2000, A\&A 356, 590

\rf{} van den Ancker M.E., Perez M.R., de Winter D., McCollum B., 2000, A\&A 363, L25

\rf{} Worley C.E., Douglass G.G., 1996, The Washington Visual Double Star Catalog (WDS)

\rf{} Wuchterl G. \& Tschanuter W.M., 2003, A\&A 398, 1081

\rf{} Zucker S. \& Mazeh T., 2002, ApJ 562, 1038

\rf{} Zuckerman B., Webb R.A., 2000, ApJ 535, 959 (ZW00)

\rf{} Zuckerman B., Song I., Webb R.A., 2001a, ApJ 559, 388

\rf{} Zuckerman B., Song I., Bessell M.S., Webb R.A., 2001b, ApJ 562, L87 (Z01)

\abe

\end{document}